# Investigating the properties of the near contact binary system TW CrB

G. Faillace, C. Owen, D. Pulley & D. Smith

## Abstract

TW Coronae Borealis (TW CrB) is a binary system likely to be active showing evidence of starspots and a hotspot. We calculated a new ephemeris based on all available timings from 1946 and find the period to be 0.58887492(2) days. We have revised the average rate of change of period down from $1.54(16) \times 10^{-7}$ days $yr^{-1}$ to $6.66(14) \times 10^{-8}$ days $yr^{-1}$. Based on light curve simulation analysis we conclude that the two stars are close to filling their Roche lobes, or possibly that one of the stars' Roche lobe has been filled. The modelling also led to a hotspot and two starspots being identified. Conservative mass transfer is one of a number of possible mechanisms considered that could explain the change in period, but the mass transfer rate would be significantly lower than previous estimates. We found evidence that suggests the period was changing in a cyclical manner, but we do not have sufficient data to make a judgement on the mechanism causing this variation. The existence of a hotspot suggests mass transfer with a corresponding increase in the amplitude in the B band as compared with the R and V bands. The chromospheric activity implied by the starspots make this binary a very likely X-ray source.

## Introduction

Close binary systems are classified according to the shape of their combined light curve, as well as the physical and evolutionary characteristics of their components. Key factors in their classification are the position of the components relative to the barycentre of the system, their colour, magnitude and in particular the degree to which their Roche lobes have been filled. Eclipsing binaries have orbital planes close to the observer's line of sight so that components eclipse each other with a consequent periodic change in the brightness of the system. TW CrB is such an eclipsing binary, its short period of less than one day, plus the degree to which the components fill their respective Roche lobes, results in a classification of near contact binary that may possess close evolutionary connections with the W UMa systems.[1] In these systems the onset of eclipses are difficult to pinpoint exactly from their light curves due to the component stars ellipsoidal shapes resulting from their mutual strong gravitational interaction.

TW CrB has been studied since 1946 but to our knowledge no spectroscopic radial velocity measurements have been published. The first ephemeris for TW CrB was compiled in 1973 by V. P. Tsesevich of the Astronomical Observatory, Odessa State University, Ukraine based upon fourteen data sets. Extensive data have been subsequently published by BBSAG (the Swiss Astronomical Society) and other authors leading to a review of this binary system by Zhang & Zhang in 2003. [1] The analysis by Zhang & Zhang spanned the period 1974 to 2001 and provided a revised ephemeris which indicated that the system is a detached near-contact binary with rapidly increasing period. Their analysis also suggested that the primary component was a slightly evolved main sequence star of spectral type F8 with an under-massive secondary.

More recently Caballero-Nieves et al. reported that their observations using multiband photometry indicated a combination of an early A0 and late K0 spectral type.[2] In addition there was a strong suggestion that one or both of the components exhibited a hot spot with at least one of these objects filling the Roche lobe. In 2010 and 2011 we performed CCD photometry on TW CrB and these times to minima, together with all known earlier timings, have allowed us to construct new light curves and re-compute the ephemeris based upon 200 data sets spanning the period 1946 to 2011. In this paper we present these observations together with a revised photometric and simulation analysis of this binary system.

**Observations**

Using our frames taken on 25[th] and 30[th] May, 2010 at the remotely operable Sierra Stars Observatory, California, we measured the position of target star TW CrB, with *Astrometrica* using the UCAC3 (US Naval Observatory Astrograph CCD, 2009) catalogue.[3] The position was measured as: RA 16h 06m 50.679±+/-0.007s, Dec +27° 16′ 34.62″ +/-± 0.08″ (2000). The Tycho catalogue lists the parallax of TW CrB as 31.1 mas, equivalent to a distance of 32.2 pc, with a proper motion (pmRA) of -27.5 mas/yr and pmDE of -4.2mas/yr.

Our observations of TW CrB were carried out during the months of May and June of 2010 and 2011. These observations were made from four different locations listed in Table 1.

Table 1. Our observations, May & June 2010/2011

| Observation Site/instrumentation | Start Time (JD) | Observing Team |
|---|---|---|
| **Observatorio Astronómico de Mallorca** | 2455329.339 | D.Smith/Faillace/Treasure/Grant |
| **07144 Costrix, Spain** | 2455332.342 | Smith. N/Grant |
| Celestron 14 telescope; | 2455335.353 | Pulley/Treasure |
| 3910mm FL @ f/11, | 2455338.353 | Owen/Pulley |
| SBIG camera STL 1001E, | 2455341.347 | Smith.N/Hajducki |
| 1024 X 1024 PIXELS @ 24µm, | 2455344.426 | Smith. D/Pulley |
| 22 x 22 arc min field of view. | 2455347.400 | Faillace/Pulley |
| | 2455350.380 | Owen/Treasure |
| | 2455353.366 | Faillace/Cornwall |
| **Sierra Stars Observatory, Markleeville,** | 2455323.791 | Faillace |
| **California, USA** | 2455337.732 | Faillace |
| Nighthawk CC06 telescope; | 2455341.777 | Faillace |
| 6100 mm FL @ f/10, | 2455346.749 | Faillace |
| Finger Lakes Instrumentation Pro Line camera, | 2455681.770 | Faillace |
| Kodak KAF-09000 3056 x 3056 pixel CCD, | 2455684.719 | Owen |
| 21 x 21 arcmin field of view. | 2455705.928 | Pulley |
| | 2455708.873 | Owen |
| | 2455737.728 | Pulley |
| | 2455738.906 | Smith. D/Pulley |
| **GRAS - New Mexico, USA** | | |
| Deep Space - Takahashi Mewlon telescope, | 2455328.858 | Faillace |
| 3572mm FL @ f/11.9, | 2455336.894 | Faillace |
| Main Camera: FLI IMG 1024 Dream Machine | 2455339.712 | Faillace |
| 1024 X 1024 PIXELS @ 24µm, | | |
| 23.6 x 23.6 arcmin field of view. | | |
| **South Stoke, United Kingdom** | 2455739.451 | Faillace |
| 11" SCT f/5.0 Camera: SBIG ST 9EXE | 2455742.411 | Faillace |

Throughout our work all recorded times were corrected to heliocentric Julian dates (HJD). Early observing sessions concentrated on capturing as much information from the target as possible using Johnson B, V and R filters. Analysis of the data from these early sessions was then used to plan later observing sessions. At the start of every observing session we took dusk flats using the filters scheduled for use in the evening's observing session. Bias frames and dark frames covering the exposures planned for the session were also taken. Frames were selected at random for assessing image quality which included checking saturation levels, monitoring SNR ratios, looking for signs of star trails and other irregularities.

**Photometric Analysis**
*Reference Stars and Differential Photometry*
We carried out our initial photometric analysis using the MaxIm DL software package employing differential aperture photometry to generate light curves for each band in order to determine the system's minima.[4]   Using the Aladin Sky Atlas,[5] a star known to be of constant magnitude was selected as a reference star (Table 2a), which is then compared with the changing magnitude of the target star using the software package.  Other non variable stars (Table 2a) were also selected that were of similar magnitude to the target star.  These check stars were not used in the analysis, but were used to ensure that there was no variability in the reference star.  The standard deviation of the check star light curves and the SNR values of the target star were used to calculate uncertainties. The phase folded light curves obtained are shown in Figure 1.

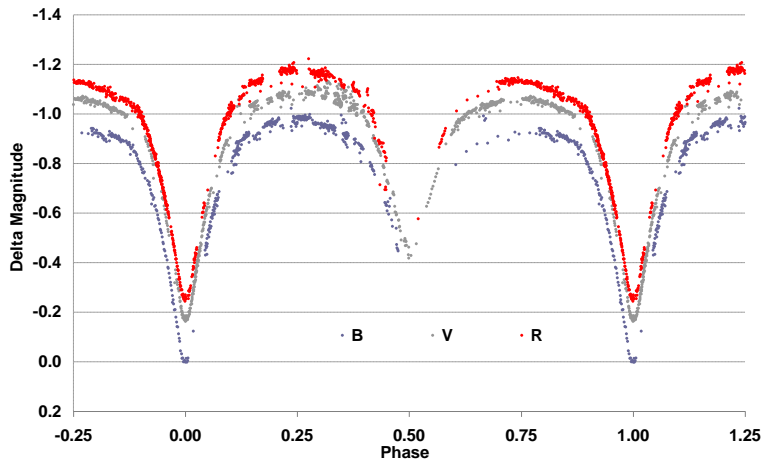

*Figure 1. TW CrB V/B/R filter phase folded light curve from data collected 7th May to 5th June 2010.  Typical uncertainties in the measurements lie in the range +/- 0.03 mag for B filter; +/- 0.01 for V filter; +/- 0.02 for R filter.  Phase and B filter minima zeroed for direct comparison of light curves of different filters.*

Table 2a. Coordinates of  stars used in MaxIm DL differential photometry

| Photometry Identity | Catalog Number | RA J2000 (2MASS) | DEC J2000 (2MASS) |
| --- | --- | --- | --- |
| Target Star | TYC 2038-1478-1 | 16 06 50.703 | +27 16 34.58 |
| Reference Star | TYC 2038-1347-1 | 16 06 25.203 | +27 18 29.97 |
| Check Star | GSC 02038-01473 | 16 06 23.262 | +27 17 16.64 |
| Check Star | GSC 02038-01381 | 16 06 38.770 | +27 16 16.32 |
| Check Star | GSC 02038-01346 | 16 07 15.050 | +27 11 26.31 |
| Check Star | GSC 02038-01353 | 16 07 12.366 | +27 20 00.15 |

Catalogue-based magnitudes of TW CrB were derived using the photometry software package Canopus.[6] The results of this analysis were used to generate phase and normalised flux values compatible with the Binary Maker 3 system modelling programme.[7] Canopus V10 makes available the magnitudes taken from the Carlsberg Meridian Catalogue 14 (CMC 14) and the Sloan Digital Sky Survey (SDSS) catalogue and transforms the J-K magnitude of 2 MASS and the r' magnitude of SDSS to BVRI magnitudes. The values obtained from these sources are consistent to within 0.02 magnitudes when using a calibration method involving the Canopus add-on, Comparison Star Selector, which picks comparison stars of similar colour and magnitude to the target star. Four comparison stars were used for this photometry and the same comparison stars were used for each image.  Table 2b shows the location of these comparison stars and their magnitude. Using the average derived magnitudes of the target star from each comparison star and the standard deviation of the average, the final value for the target star was obtained for each frame. The standard deviation incorporates the uncertainty in the measurement of the target and comparison stars, taken from the SNR values of the target and the reference star, the uncertainty in the catalogue value and the uncertainty in the correction for colour difference. In order to reduce scatter and to enable smoothing of the phased light curves, adjacent data points up to a maximum interval of four minutes were binned and averages computed. A normalised flux light curve was then obtained for each of the three bands using a Fourier transform fit. This binning had virtually no effect on the Fourier coefficients (to the third decimal place) nor to the normalised phased and flux values used in Binary Maker 3 modelling.

Table 2b. Coordinates and derived B, V and R magnitudes of comparison stars used for absolute photometry in the Canopus software package.

| Catalog Number | RA J2000 (2MASS) | DEC J2000 (2MASS) | B | V | R | r' | B-V | V-R |
|---|---|---|---|---|---|---|---|---|
| GSC 02038-01473 | 16 06 23.262 | +27 17 16.64 | 13.46 | 12.75 | 12.35 | 12.58 | 0.71 | 0.40 |
| GSC 02038-01577 | 16 06 27.648 | +27 17 04.25 | 14.64 | 13.90 | 13.48 | 13.69 | 0.74 | 0.42 |
| GSC 02038-01270 | 16 07 07.748 | +27 18 00.06 | 13.69 | 13.16 | 12.85 | 13.05 | 0.54 | 0.31 |
| GSC 02038-01672 | 16 07 07.170 | +27 20 04.75 | 15.53 | 15.00 | 14.68 | 14.99 | 0.54 | 0.31 |
| | | Average: | 14.33 | 13.70 | 13.34 | 13.58 | 0.63 | 0.36 |

### Qualitative Light Curve Analysis

Visual inspection of the light curve provided basic information about this binary system. We conclude that: (i) the system does not have a relatively flat "out of eclipse" light curve; (ii) the system shows significant differences in eclipse depths indicating differences in temperatures for the two components; (iii) the light curves vary continually, even when not eclipsing, because the components visible cross-sectional areas are continually changing. This implies that the surfaces of these two components are in close proximity to the critical Roche equipotential containing the inner Lagrangian point, and their shapes are being distorted into ellipsoids because of gravitational and tidal forces; (iv) at minima, the light curves are not flattened so the eclipses are not total; (v) the short orbital period of the system, less than 0.6 days, is indicative of their close proximity and (vi) in the combined band phase plots, Figure 1, it can be seen that the system becomes redder during the eclipses as can be seen by the B light curve fading more than the V and R light curves.

We can conclude from the light curve that the two stars are unlikely to be in contact, or over contact, although the possibility cannot be ruled out.

Visual inspection of the light curves, once data misalignment has been excluded, indicates the presence of the O'Connell effect whereby the two out of eclipse maxima of the light curves are unequal. [8] [9] The maxima are expected to be equally high, because the observed luminosity of an eclipsing binary system when the two components are side by side should be equal to the luminosity in the configuration half an orbital period later when they have switched positions. The O'Connell effect is an area of ongoing research. This particular feature was not reported by earlier observers which suggests that this is an active system that may have been in a period of quiescence.[1] [2]

### Period Analysis

All known times of minima for TW CrB are listed in Table 3. This includes the 112 timings contained in the Zhang & Zhang analysis,[1] and a further 91 timings taken from other related sources identified in Table 3. The data sets include 12 times of minimum obtained from our CCD measurements using The Open University's remotely-operable PIRATE facility (2010),[10] and Sierra Stars Observatory, California (2010 and 2011). These timings were analysed using the Kwee-van Woerden methodology,[11] contained within the Peranso period analysis software.[12] All pre 1974 timings are from photographic imaging and post 1998 timings are from either photoelectric (pe) or CCD imaging. With the exception of one photoelectric observation (E = -3,058.5) all timings recorded between 1974 and 1998 are visual. The timings of minima span 65 years permitting a detailed investigation into the long term variation in period of TW CrB. When calculating the ephemerides we have assigned weightings of 1 to visual minima; 2 to photographic minima and 10 to CCD and pe minima. This was in line with the weighting applied by Zhang & Zhang.[1]

Linear and second order polynomial regression analysis was applied to the 203 timings to generate new linear and quadratic ephemerides. The plot of the O − C residuals led us to eliminate three data sets (E = -2,335.0; -2,323.0; +1,484.0) whose O − C values deviated by more than 4 sigma, for their observing methodology, from the quadratic fit. The resulting ephemerides, with standard errors in parenthesis, calculated from the remaining 200 timings are:

HJD Min$_{lin}$   = 2,451,273.4740(2) + 0.58887492(2)E                (1)
HJD Min$_{quad}$ = 2,451,273.4701(1) + 0.58887562(2)E + 5.37(11)x10$^{-11}$E$^2$   (2)

Our calculated linear O - C residuals are listed in Table 3 and displayed in Figure 2 together with the quadratic curve for the elements above. This curve suggests that the period of TW CrB is increasing with time and consistent with secular mass transfer between the binary components. The average rate of period increase, taken from the quadratic ephemeris, equates to 1.07(2) x 10$^{-10}$ days per cycle or 6.66(14) x 10$^{-8}$ days yr$^{-1}$.

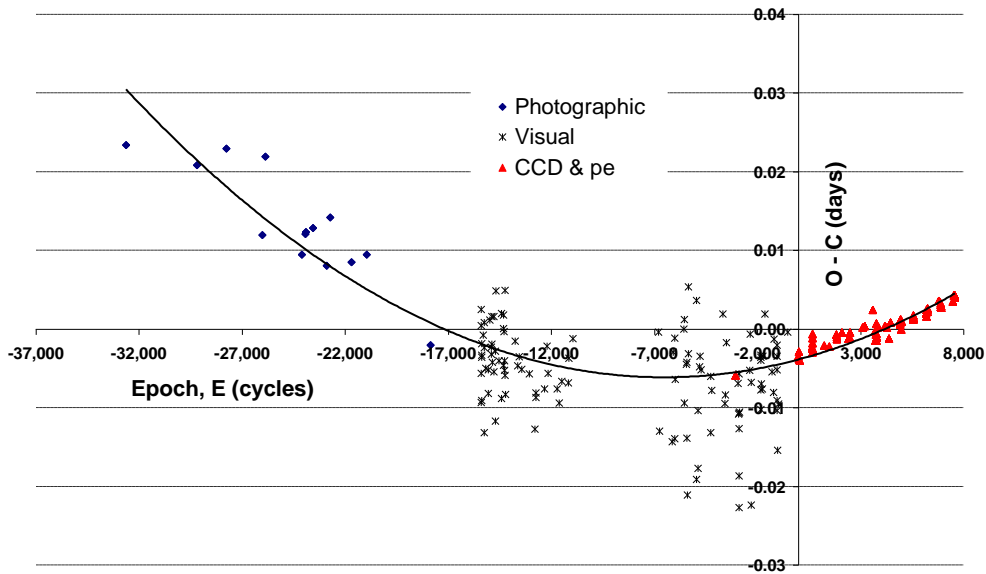

*Figure 2.  O - C plot for TW CrB from 1946 to 2011 and displaying the quadratic fit to the residuals*

Table 3.  Times of minima of TW CrB

| HJD (2,400,000 +) | Meth | E | O - C | Ref | HJD (2,400,000 +) | Meth | E | O - C | Ref |
|---|---|---|---|---|---|---|---|---|---|
| 32,061.4530 | photo | -32,625.0 | 0.0234 | 1 | 47,754.3510 | vis | -5,976.0 | -0.0064 | 1, 2, 8 |
| 34,092.4800 | photo | -29,176.0 | 0.0208 | 1 | 47,996.3850 | vis | -5,565.0 | 0.0000 | 1, 2, 8 |
| 34,926.3290 | photo | -27,760.0 | 0.0229 | 1 | 48,013.4530 | vis | -5,536.0 | -0.0094 | 1, 2, 8 |
| 35,957.4380 | photo | -26,009.0 | 0.0119 | 1 | 48,016.4080 | vis | -5,531.0 | 0.0012 | 1, 2, 8 |
| 36,037.5350 | photo | -25,873.0 | 0.0219 | 1 | 48,069.4010 | vis | -5,441.0 | -0.0045 | 1, 2, 8 |
| 37,080.4200 | photo | -24,102.0 | 0.0094 | 1 | 48,086.4690 | vis | -5,412.0 | -0.0139 | 1, 2, 8 |
| 37,191.7200 | photo | -23,913.0 | 0.0121 | 1, 8 | 48,089.4230 | vis | -5,407.0 | -0.0043 | 1, 2, 8 |
| 37,202.3200 | photo | -23,895.0 | 0.0123 | 1, 8 | 48,099.4170 | vis | -5,390.0 | -0.0211 | 1, 2, 8 |
| 37,402.5380 | photo | -23,555.0 | 0.0128 | 1, 8 | 48,125.3540 | vis | -5,346.0 | 0.0054 | 1, 2, 8 |
| 37,789.4240 | photo | -22,898.0 | 0.0080 | 1, 8 | 48,132.4120 | vis | -5,334.0 | -0.0031 | 1, 2, 8 |
| 37,898.3720 | photo | -22,713.0 | 0.0142 | 1, 8 | 48,357.3690 | vis | -4,952.0 | 0.0036 | 1, 2, 8 |
| 38,502.5520 | photo | -21,687.0 | 0.0085 | 1, 8 | 48,358.5240 | vis | -4,950.0 | -0.0191 | 1, 2, 8 |
| 38,935.3760 | photo | -20,952.0 | 0.0094 | 1, 8 | 48,404.4650 | vis | -4,872.0 | -0.0103 | 1, 2, 8 |
| 40,764.4100 | photo | -17,846.0 | -0.0021 | 1, 8 | 48,407.4020 | vis | -4,867.0 | -0.0177 | 1, 2, 8 |
| 42,200.3840 | vis | -15,407.5 | 0.0004 | 1, 2, 8 | 48,442.4530 | vis | -4,807.5 | -0.0048 | 1, 2, 8 |
| 42,201.5520 | vis | -15,405.5 | -0.0093 | 1, 2, 8 | 48,460.4150 | vis | -4,777.0 | -0.0035 | 1, 2, 8 |
| 42,202.4390 | vis | -15,404.0 | -0.0057 | 1, 2, 8 | 48,480.4350 | vis | -4,743.0 | -0.0052 | 1, 2, 8 |
| 42,212.4580 | vis | -15,387.0 | 0.0025 | 1, 2, 8 | 48,758.3760 | vis | -4,271.0 | -0.0132 | 1, 2, 8 |
| 42,214.5130 | vis | -15,383.5 | -0.0036 | 1, 2, 8 | 48,768.3940 | vis | -4,254.0 | -0.0061 | 1, 2, 8 |
| 42,215.3980 | vis | -15,382.0 | -0.0019 | 1, 2, 8 | 48,788.4140 | vis | -4,220.0 | -0.0078 | 1, 2, 8 |
| 42,220.4020 | vis | -15,373.5 | -0.0033 | 1, 2, 8 | 49,116.4270 | vis | -3,663.0 | 0.0019 | 1, 2, 8 |
| 42,221.5740 | vis | -15,371.5 | -0.0091 | 1, 2, 8 | 49,166.4700 | vis | -3,578.0 | -0.0095 | 1, 2, 8 |
| 42,258.3870 | vis | -15,309.0 | -0.0008 | 1, 2, 8 | 49,176.4820 | vis | -3,561.0 | -0.0084 | 1, 2, 8 |
| 42,288.4180 | vis | -15,258.0 | -0.0024 | 1, 2, 8 | 49,212.4100 | vis | -3,500.0 | -0.0017 | 1, 2, 8 |
| 42,296.3570 | vis | -15,244.5 | -0.0132 | 1, 2, 8 | 49,472.3941 | pe | -3,058.5 | -0.0059 | 1, 2, 8 |
| 42,296.3710 | vis | -15,244.5 | 0.0008 | 1, 2, 8 | 49,550.4190 | vis | -2,926.0 | -0.0070 | 1, 2, 8 |
| 42,337.2920 | vis | -15,175.0 | -0.0050 | 1, 2, 8 | 49,560.4260 | vis | -2,909.0 | -0.0108 | 1, 2, 8 |
| 42,404.7150 | vis | -15,060.5 | -0.0082 | 1, 2, 8 | 49,570.4250 | vis | -2,892.0 | -0.0227 | 2, 8 |
| 42,404.7180 | vis | -15,060.5 | -0.0052 | 1, 2, 8 | 49,570.4330 | pe | -2,892.0 | -0.0187 | 2, 8 |
| 42,455.6620 | vis | -14,974.0 | 0.0011 | 1, 2, 8 | 49,570.4350 | vis | -2,892.0 | -0.0127 | 1, 2, 8 |
| 42,491.5790 | vis | -14,913.0 | -0.0032 | 1, 2, 8 | 49,570.4370 | vis | -2,892.0 | -0.0107 | 2, 8 |
| 42,493.6410 | vis | -14,909.5 | -0.0023 | 1, 2, 8 | 49,570.4420 | vis | -2,892.0 | -0.0057 | 2, 8 |
| 42,509.5410 | vis | -14,882.5 | -0.0019 | 1, 2, 8 | 49,580.4480 | vis | -2,875.0 | -0.0106 | 1, 2, 8 |
| 42,516.6110 | vis | -14,870.5 | 0.0016 | 1, 2, 8 | 49,878.4240 | vis | -2,369.0 | -0.0053 | 1, 2, 8 |
| 42,524.5550 | vis | -14,857.0 | -0.0042 | 1, 2, 8 | 49,895.5060 | vis | -2,340.0 | -0.0007 | 1, 2, 8 |
| 42,568.4320 | vis | -14,782.5 | 0.0016 | 1, 2, 8 | *49,898.4187* | vis | *-2,335.0* | *-0.0323* | *8* |
| 42,570.4860 | vis | -14,779.0 | -0.0055 | 1, 2, 8 | *49,905.4864* | vis | *-2,323.0* | *-0.0311* | *8* |
| 42,606.4100 | vis | -14,718.0 | -0.0029 | 1, 2, 8 | 49,915.5060 | vis | -2,306.0 | -0.0224 | 1, 8 |
| 42,616.4120 | vis | -14,701.0 | -0.0117 | 1, 2, 8 | 49,918.4660 | vis | -2,301.0 | -0.0068 | 1, 2, 8 |
| 42,621.4340 | vis | -14,692.5 | 0.0048 | 1, 2, 8 | 50,189.3450 | vis | -1,841.0 | -0.0102 | 1, 8 |
| 42,716.2340 | vis | -14,531.5 | -0.0040 | 1, 2, 8 | 50,193.4696 | vis | -1,834.0 | -0.0078 | 1, 8 |
| 42,780.7110 | vis | -14,422.0 | -0.0088 | 1, 2, 8 | 50,193.4738 | vis | -1,834.0 | -0.0036 | 1, 8 |
| 42,791.6160 | vis | -14,403.5 | 0.0020 | 1, 2, 8 | 50,193.4740 | vis | -1,834.0 | -0.0034 | 10 |
| 42,836.6630 | vis | -14,327.0 | 0.0001 | 1, 2, 8 | 50,200.5399 | vis | -1,822.0 | -0.0040 | 1, 8 |
| 42,837.5480 | vis | -14,325.5 | 0.0017 | 1, 2, 8 | 50,209.3700 | vis | -1,807.0 | -0.0070 | 1, 2, 8 |
| 42,840.4860 | vis | -14,320.5 | -0.0046 | 1, 2, 8 | 50,239.4020 | vis | -1,756.0 | -0.0076 | 1, 2, 8 |
| 42,858.4510 | vis | -14,290.0 | -0.0003 | 1, 2, 8 | 50,249.4170 | vis | -1,739.0 | -0.0035 | 1, 2, 8 |
| 42,870.5180 | vis | -14,269.5 | -0.0053 | 1, 2, 8 | 50,312.4320 | vis | -1,632.0 | 0.0019 | 1, 2, 8 |
| 42,878.4690 | vis | -14,256.0 | -0.0041 | 1, 2, 8 | 50,515.5865 | vis | -1,287.0 | -0.0054 | 1, 8 |
| 42,882.6000 | vis | -14,249.0 | 0.0048 | 1, 2, 8 | 50,557.3940 | vis | -1,216.0 | -0.0081 | 1, 2, 8 |
| 42,886.4170 | vis | -14,242.5 | -0.0059 | 1, 2, 8 | 50,570.3560 | vis | -1,194.0 | -0.0013 | 1, 2, 8 |
| 42,905.5530 | vis | -14,210.0 | -0.0083 | 1, 2, 8 | 50,660.4448 | vis | -1,041.0 | -0.0104 | 1, 8 |
| 43,177.6200 | vis | -13,748.0 | -0.0015 | 1, 2, 8 | 50,660.4461 | vis | -1,041.0 | -0.0091 | 1, 8 |
| 43,254.4650 | vis | -13,617.5 | -0.0047 | 1, 2, 8 | 50,660.4500 | vis | -1,041.0 | -0.0052 | 1, 2, 8 |
| 43,295.3930 | vis | -13,548.0 | -0.0035 | 1, 2, 8 | 50,660.4517 | vis | -1,041.0 | -0.0035 | 1, 8 |
| 43,358.4010 | vis | -13,441.0 | -0.0051 | 1, 2, 8 | 50,660.4531 | vis | -1,041.0 | -0.0021 | 1, 8 |
| 43,581.5840 | vis | -13,062.0 | -0.0057 | 1, 2, 8 | 50,673.3950 | vis | -1,019.0 | -0.0154 | 1, 2, 8 |
| 43,734.3900 | vis | -12,802.5 | -0.0128 | 1, 2, 8 | 50,696.3670 | vis | -980.0 | -0.0096 | 1, 2, 8 |
| 43,765.3100 | vis | -12,750.0 | -0.0087 | 1, 2, 8 | 50,956.0700 | vis | -539.0 | -0.0004 | 1, 8 |
| 43,770.3160 | vis | -12,741.5 | -0.0081 | 1, 2, 8 | 51,273.4705 | pe | 0.0 | -0.0035 | 1, 3, 8 |
| 44,022.3550 | vis | -12,313.5 | -0.0076 | 1, 2, 8 | 51,274.0590 | ccd | 1.0 | -0.0039 | 10 |
| 44,085.3700 | vis | -12,206.5 | -0.0022 | 1, 2, 8 | 51,283.7764 | ccd | 17.5 | -0.0029 | 1, 4, 8 |
| 44,123.3490 | vis | -12,142.0 | -0.0057 | 1, 2, 8 | 51,311.7468 | ccd | 65.0 | -0.0040 | 1, 4, 8 |
| 44,382.4520 | vis | -11,702.0 | -0.0076 | 1, 2, 8 | 51,659.4790 | pe | 655.5 | -0.0025 | 1, 8 |
| 44,437.5100 | vis | -11,608.5 | -0.0094 | 1, 2, 8 | 51,664.1902 | ccd | 663.5 | -0.0023 | 1, 5 |
| 44,502.3890 | vis | -11,498.5 | -0.0067 | 1, 2, 8 | 51,665.0746 | ccd | 665.0 | -0.0012 | 1, 5 |
| 44,701.6230 | vis | -11,160.0 | -0.0068 | 1, 2, 8 | 51,671.2583 | ccd | 675.5 | -0.0007 | 1, 5 |
| 44,711.6370 | vis | -11,143.0 | -0.0037 | 1, 2, 8 | 51,672.1405 | ccd | 677.0 | -0.0018 | 1, 5 |
| 44,824.4090 | vis | -10,951.5 | -0.0013 | 1, 2, 8 | 51,675.0837 | ccd | 682.0 | -0.0030 | 1, 5 |
| 47,274.4240 | vis | -6,791.0 | -0.0004 | 1, 2, 8 | 51,680.3839 | pe | 691.0 | -0.0026 | 1, 8, 9 |
| 47,304.4440 | vis | -6,740.0 | -0.0130 | 1, 2, 8 | 52,009.5655 | ccd | 1,250.0 | -0.0021 | 1, 8, 9 |
| 47,665.4230 | vis | -6,127.0 | -0.0143 | 1, 2, 8 | 52,147.3621 | ccd | 1,484.0 | -0.0023 | 1, 8, 9 |
| 47,695.4670 | vis | -6,076.0 | -0.0029 | 1, 2, 8 | *52,147.3747* | ccd | *1,484.0* | *0.0103* | *1, 8, 9* |
| 47,728.4330 | vis | -6,020.0 | -0.0139 | 1, 2, 8 | 52,352.2921 | pe | 1,832.0 | -0.0007 | 1, 8 |
| 47,741.4010 | vis | -5,998.0 | -0.0012 | 1, 2, 8 | 52,360.5358 | pe | 1,846.0 | -0.0013 | 1, 8 |

Table 3 contd.  Times of minima of TW CrB

| HJD (2,400,000 +) | Meth | E | O - C | Ref | HJD (2,400,000 +) | Meth | E | O - C | Ref |
|---|---|---|---|---|---|---|---|---|---|
| 52,373.4913 | pe | 1,868.0 | -0.0010 | 1, 8, 9 | 54,213.4330 | ccd | 4,992.5 | 0.0010 | 8 |
| 52,510.4053 | ccd | 2,100.5 | -0.0005 | 1, 8, 9 | 54,556.4529 | ccd | 5,575.0 | 0.0012 | 8 |
| 52,692.6613 | ccd | 2,410.0 | -0.0012 | 1, 8, 9 | 54,556.4530 | ccd | 5,575.0 | 0.0013 | 8 |
| 52,721.5164 | pe | 2,459.0 | -0.0010 | 1, 8, 9 | 54,556.4531 | ccd | 5,575.0 | 0.0014 | 8 |
| 52,741.5387 | pe | 2,493.0 | -0.0005 | 1, 8, 9 | 54,556.4534 | | 5,575.0 | 0.0017 | 10 |
| 53,107.5251 | pe | 3,114.5 | 0.0002 | 1, 8, 9 | 54,911.5452 | ccd | 6,178.0 | 0.0019 | 8 |
| 53,165.5295 | pe | 3,213.0 | 0.0004 | 1, 8, 9 | 54,924.5009 | ccd | 6,200.0 | 0.0024 | 8 |
| 53,388.7151 | ccd | 3,592.0 | 0.0024 | 1, 8, 9 | 54,930.3888 | ccd | 6,210.0 | 0.0015 | 8 |
| 53,448.1887 | ccd | 3,693.0 | -0.0004 | 1, 8, 9 | 54,950.4115 | ccd | 6,244.0 | 0.0025 | 8 |
| 53,463.4983 | pe | 3,719.0 | -0.0015 | 1, 8, 9 | 54,950.4116 | ccd | 6,244.0 | 0.0026 | 8 |
| 53,473.8044 | ccd | 3,736.5 | -0.0007 | 1, 8, 9 | 54,960.1276 | ccd | 6,260.5 | 0.0022 | 8 |
| 53,492.3532 | ccd | 3,768.0 | -0.0015 | 1, 8 | 55,269.8772 | ccd | 6,786.5 | 0.0036 | 8 |
| 53,493.5331 | ccd | 3,770.0 | 0.0007 | 1, 8 | 55,293.4322 | ccd | 6,826.5 | 0.0036 | 8 |
| 53,499.4206 | ccd | 3,780.0 | -0.0006 | 1, 8 | 55,332.5919 | ccd | 6,893.0 | 0.0031 | 7 |
| 53,502.3644 | ccd | 3,785.0 | -0.0012 | 1, 8 | 55,335.5364 | ccd | 6,898.0 | 0.0032 | 7 |
| 53,503.5423 | pe | 3,787.0 | -0.0010 | 1, 8, 9 | 55,338.4804 | ccd | 6,903.0 | 0.0028 | 7 |
| 53,550.0647 | pe | 3,866.0 | 0.0003 | 1, 8 | 55,341.4246 | ccd | 6,908.0 | 0.0026 | 7 |
| 53,747.3379 | pe | 4,201.0 | 0.0004 | 1, 8 | 55,349.0802 | | 6,921.0 | 0.0029 | 10 |
| 53,801.5142 | ccd | 4,293.0 | 0.0002 | 1, 8 | 55,681.7951 | ccd | 7,486.0 | 0.0034 | 7 |
| 53,859.5170 | pe | 4,391.5 | -0.0012 | 1, 8 | 55,684.7395 | ccd | 7,491.0 | 0.0035 | 7 |
| 53,900.4458 | ccd | 4,461.0 | 0.0008 | 1, 8 | 55,705.9397 | ccd | 7,527.0 | 0.0042 | 7 |
| 54,172.5058 | ccd | 4,923.0 | 0.0006 | 1, 8 | 55,708.8842 | ccd | 7,532.0 | 0.0043 | 7 |
| 54,172.5060 | ccd | 4,923.0 | 0.0008 | 8 | 55,737.7390 | ccd | 7,581.0 | 0.0042 | 7 |
| 54,172.5063 | ccd | 4,923.0 | 0.0011 | 8 | 55,738.9165 | ccd | 7,583.0 | 0.0040 | 7 |
| 54,185.4615 | ccd | 4,945.0 | 0.0010 | 8 | 55,739.5053 | ccd | 7,584.0 | 0.0039 | 7 |
| 54,185.4617 | ccd | 4,945.0 | 0.0012 | 1, 8 | 55,742.4501 | ccd | 7,589.0 | 0.0043 | 7 |
| 54,199.5934 | pe | 4,969.0 | -0.0001 | 8 | | | | | |

Note: The values highlighted were not used to calculate the ephemerides.
Vis ~ visual; pe ~ photoelectric; photo ~ photographic
References: (1) Krakow [http://www.as.up.krakow.pl/o-c/]; (2) BBSAG Bulletins [http://www.astroinfo.ch/bbsag/bbsag_e.html] (3) Agerer & Hubscher 2000; (4) Diethelm 2001; (5) Zhang & Zhang 2003; (6) Baldinelli & Maitan 2002; (7)This study; (8) Lichtenknecker BAV; (9) IBVS [http://www.konkoly.hu/IBVS/IBVS.html]; (10) GSV Search Gateway; Czech Astronomical Society

## Light Curve Simulation

We simulated the observed lightcurves using the software package Binary Maker 3.[7] This program is similar to, but not a derivative of, the Wilson-Devinney code;[13] it uses the same nomenclature but does not have the same error handling capability.  To compensate for this we used a range of values in our simulation. The programme allows the adjustment of a number of system parameters and then allows comparison of the calculated lightcurve with the observed lightcurve.

$T1_{eff}$ was determined to be 5700K±200K. This was derived from Allen's Astrophysical Quantities using the (J-K) = 0.41 derived from the 2MASS catalogue entry for TW CrB.[14]  This suggests a spectral type for TW CrB components of early G and early K.   All other parameter values were derived by minimising the difference between the observed lightcurve and the lightcurve calculated with Binary Maker 3.

Since V was the middle passband of those in which we observed TW CrB we used this band to obtain the non-wavelength dependent parameters. We then adjusted only the wavelength dependent parameters to obtain a match with the R and B passband lightcurves.

As $T1_{eff}$ is 5700K±200K this means that both $T1_{eff}$ and $T2_{eff}$ are well below 7500K and implies that both components are fully convective, hence we are justified in setting the gravity brightening coefficients g1 and g2 to 0.32.  Similarly the reflection coefficients Alb1 and Alb2 can be set to 0.5.  The limb darkening coefficients were derived from the Van Hamme table. [15]. Using this information $T2_{eff}$ was determined to be 5400K±150K.

Not having the spectroscopic data on this system needed to fix the mass ratio, q, it was estimated by minimising mass ratio over a range of values for $T1_{eff}$ and $T2_{eff}$. This led to a mass ratio of 0.725±0.010. See Fig 3.

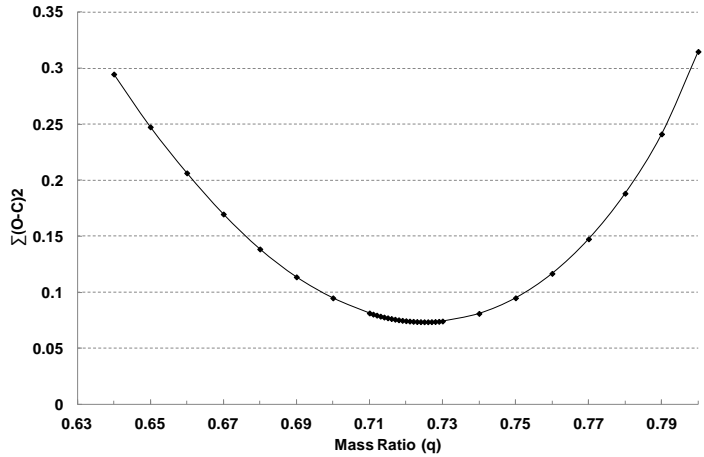

*Figure 3. Plot of the mass ratio, q, versus minimum of the $\Sigma(O - C)^2$*

In order to achieve a best fit it was necessary to set the fillout factors to -0.03281 and -0.0252, respectively. These factors indicate the degree to which the stars' physical surfaces are inside their Langrangian surfaces. These numbers are small and negative indicating that both stars are nearly filling their Roche Lobes.

Visual inspection of the lightcurves indicates that the amplitude of the B lightcurve is greater than either the R or V. This preponderance was proven in statistical analysis performed by the authors but not presented here. Figure 4 shows the observed and calculated lightcurve for the B filter.

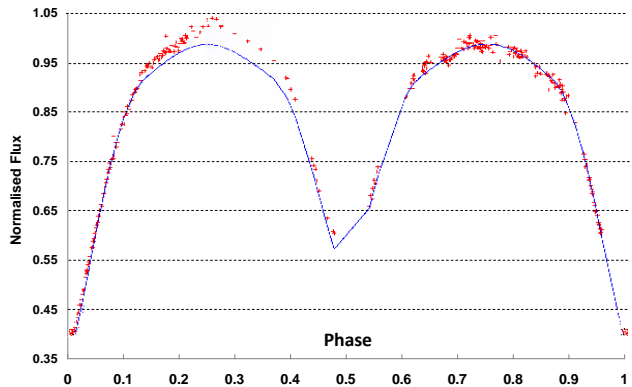

*Figure 4. B filter lightcurve with no starspots or hotspots*

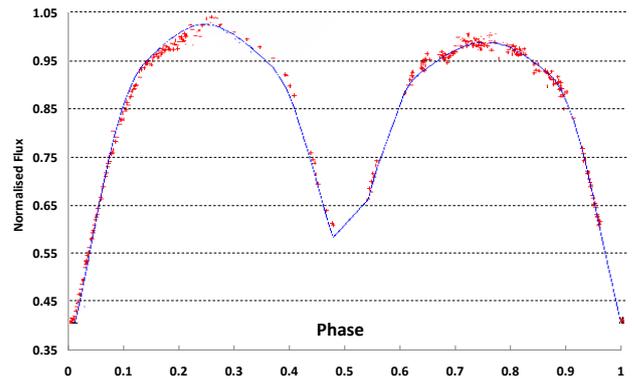

*Figure 5. B filter lightcurve with starspots or hotspots*

The increase in amplitude of the blue lightcurve, taken with the secondary being close to filling its Roche Lobe led to the introduction of a hot spot on the primary. This is consistent with the allusion to mass transfer by Zhang & Zhang and the conclusions reached by Caballero-Nieves et al who noted the same phenomena in their colour difference analysis of the system. [1][2]

In order to further improve the fit on the lightcurve shoulders it was necessary to introduce two starspots on the secondary and the combined effect can be seen in Figure 5 when compared with Figure 4. Figs 6 and 7 show the lightcurves, with starspots and hotspot, for the V and R passbands. This is shown graphically in Fig. 8.

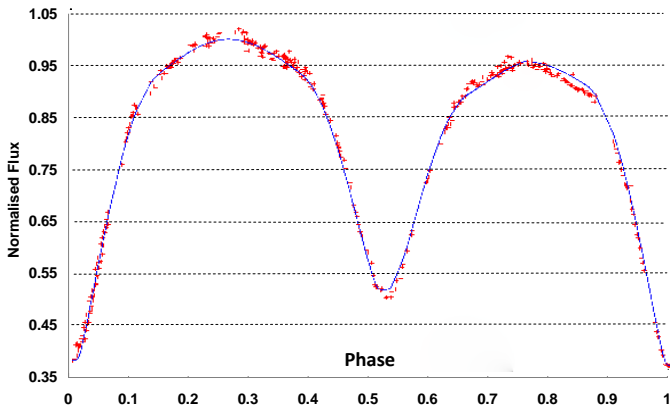

*Figure 6  V filter lightcurve with starspots or hotspots*

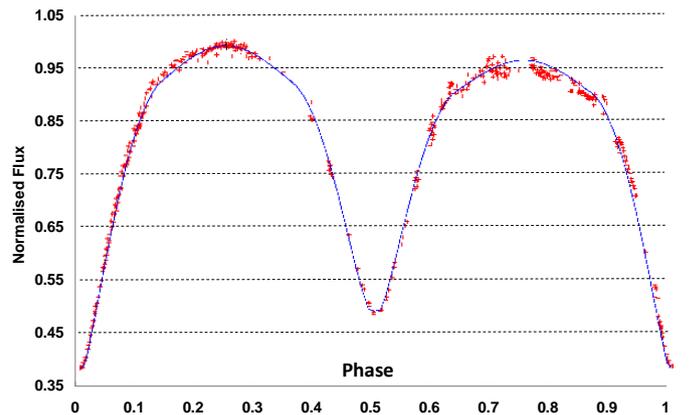

*Figure 7  R filter lightcurve with starspots or hotspots*

Table 4 lists the derived parameters of the system.

Table 4.  BM3 Derived Parameters

| Wavelength Independent Parameters | |
|---|---|
| Mass Ratio (q)$M_2/M_1$ | 0.725 ± 0.010 |
| Inclination | 89.6 ± 0.1 |
| T1$_{eff}$  (*) | 5700K ± 200K |
| T2$_{eff}$ | 5400K ± 150K |
| g1=g2: assumed | 0.32 |
| Alb1=Alb2: assumed | 0.5 |
| Omega1 | 3.33 ± 0.09 |
| Omega2 | 3.2575 ± 0.0075 |
| Fillout1 | -0.03281 |
| Fillout2 | -0.0252 |
| Lagrangian L1 | 0.53234 |
| Lagrangian L2 | 1.646104 |
| *Primary Hotspot* | |
|   Co-latitude(deg) | 93.5 |
|   Longitude(deg) | 92.8 |
|   Radius(deg) | 12 |
| *Secondary Starspot1* | |
|   Co-latitude(deg) | 57.1 |
|   Longitude(deg) | 309 |
|   Radius(deg) | 13 |
| *Secondary Starspot2* | |
|   Co-latitude(deg) | 82 |
|   Longitude(deg) | 310 |
|   Radius(deg) | 13 |

| Wavelength Dependent Parameters | | | |
|---|---|---|---|
| | | Filter | |
| | R | V | B |
| Centre Wavelength(Å) | 7000 | 5500 | 4450 |
| Luminosity 1 | 0.6232 | 0.6361 | 0.6481 |
| Luminosity 2 | 0.3768 | 0.3639 | 0.3519 |
| X1 (Limb Darkening) | 0.481 | 0.584 | 0.72 |
| X2 (Limb Darkening) | 0.504 | 0.618 | 0.774 |
| *Temp Factor (%)* | | | |
| Primary Hotspot | 126 | 95 | 117 |
| Secondary Starspot1 | 95 | 68 | 98 |
| Secondary Starspot2 | 94 | 69 | 99 |

(*)  Allen's Astrophysical Quantities

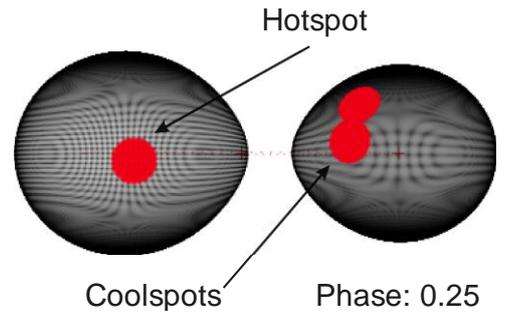

Figure 8.  Representation of TW CrB with starspots and hotspots.

## Discussion
### Orbital Period Behaviour

Examination of Figure 2 for epochs greater than zero shows that there is a systematic error with most (O − C) residuals lying above the quadratic curve for epochs up to about 4,000 and below the curve for epochs greater than 4,000. This systematic error suggests that a model of secular mass transfer between the binary components is incomplete or inappropriate for TW CrB. Possible alternative explanations considered are (i) abrupt (episodic) mass ejections or transfers; and (ii) cyclical changes possibly due to either magnetic effects or the presence of a third body.

### Abrupt (Episodic) Changes

Abrupt or episodic mass transfers can cause step changes in the orbital period of a binary system. This can be represented by a series of linear ephemerides and the corresponding (O − C) plot will consist of straight lines, each reflecting an interval of constant orbital period. Examination of Figure 2 suggests that two straight lines can be fitted to the (O − C) residuals as shown in Figure 9 with the first abrupt change occurring prior to epoch -32,061 and the second between epochs -10,951.5 and -6,791.0

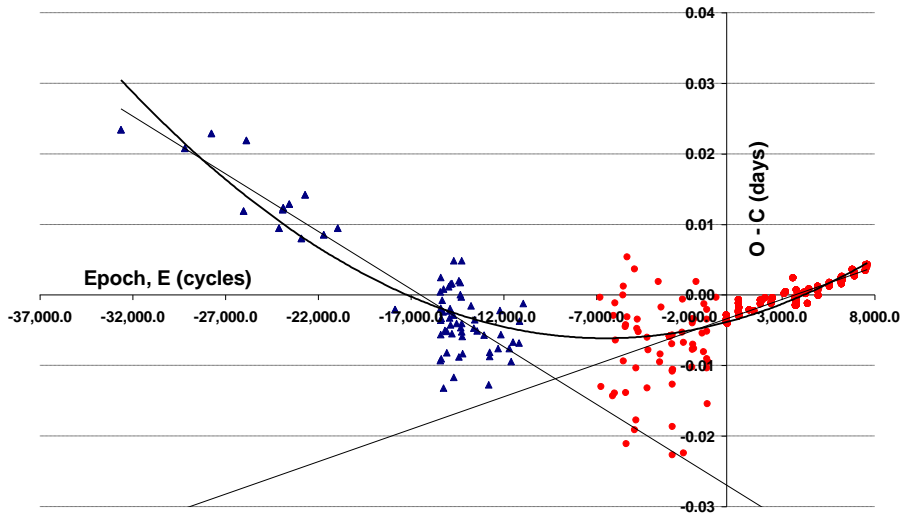

*Figure 9. O − C plot for all timings and including two linear best fit curves which could show two episodic period changes, the first prior to E = -32,625 and the second at about E = -9,500*

The corresponding linear ephemerides are:

HJD Min$_{lin1}$ = 2,451,273.4470 (1) + 0.58887329 (8)E     (for E<-10.951.5)     (3)
HJD Min$_{lin2}$ = 2, 451,273.4706 (1) + 0.58887584 (2)E   (for E>-6,791.0) (4)

However this model does not resolve all issues, particularly; (i) it does not fully explain the systematic errors in the (O − C) values seen in Figure 2 for epochs greater than zero. This is more clearly illustrated in the centre plot of Figure 10; (ii) abrupt or episodic changes would require the period to remain constant between each mass ejection. Figure 11 shows the results of an analysis of the period of TW CrB measured over 13 year intervals from 1946 which clearly shows the orbital period to be continually changing; (iii) although TW CrB is a relatively close binary system, recorded at 32.2 pc, there appears to be no supporting observational evidence for episodic mass ejections; (iv) our light curve simulation of TW CrB requires the presence of a hotspot which lends support to some form of continuous mass transfer model and not to an episodic event.

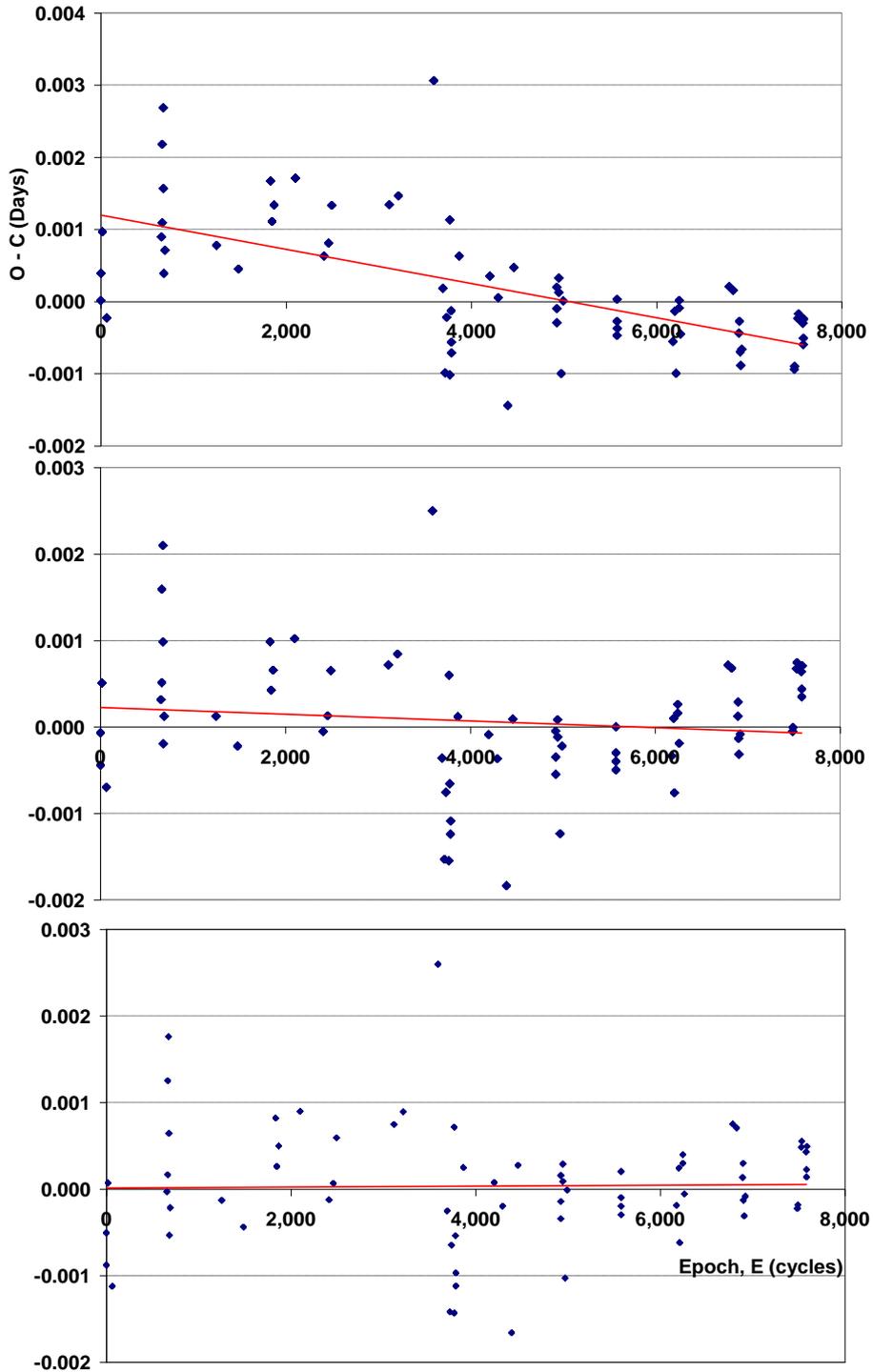

*Figure 10. Plots of the quadratic (top), linear (centre) and sine (bottom) residuals calculated from Equations (2), (4) and (5) respectively. The best fit straight line would lie along the Epoch axis.*

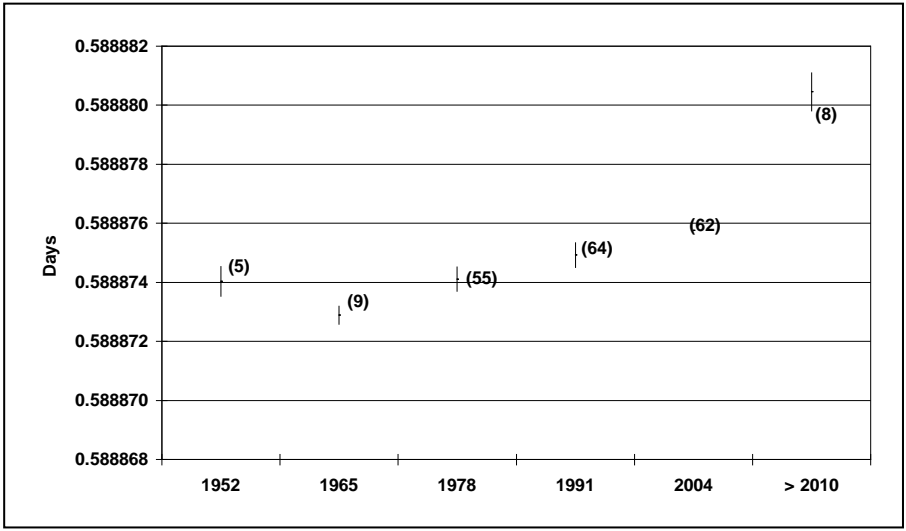

*Figure 11. The period change of TW CrB since 1946, with standard error bars and the number of timings, in each consecutive 13 year period.*

### Cyclical Changes

Some of the issues raised with the abrupt change model can be resolved by introducing a sinusoidal term into the quadratic ephemeris of Eq. (2). This takes the form of Equation (5) which combines the secular orbital period increase with a superimposed cyclical change to the orbital period:

$$HJD_{min} = A + BE + CE^2 + D \sin (\omega E + \Phi) \qquad (5)$$

Solutions to Equation (5) can be found iteratively using the Levenberg – Marquardt technique which has been implemented with OriginLab software. [16] In this implementation, the coefficients A, B and C were taken from the quadratic ephemeris of Eq. (2). These coefficients were held constant whilst varying D, $\omega$ and $\Phi$ to find the best fit to the restricted data set of E≥0. This epoch range was chosen as it makes use of only the more precise CCD and pe timings. The parameters derived from the solution to Equation (5) are listed in Table 5.

Table 5. The fitted parameters to Eq. (5)

| Parameter | Final Value | Unit |
| --- | --- | --- |
| A | 2451273.4701(1) | day |
| B | 0.58887562(2) | day |
| C | $5.37 \times 10^{-11}$(11) | day |
| D | $9.3 \times 10^{-4}$ (5) | day |
| $\omega$ | $3.6 \times 10^{-4}$(3) | rad/cycle |
| $\Phi$ | 1.3(2) | rad |
| $P_{mod}$ | 28.1(2.4) | yr |

Applegate has proposed that binary period modulation can occur when there is magnetic activity in one of the stars of a binary system.[17] Such activity can lead to changes of oblateness and angular momentum and, through gravity coupling, to orbital period modulation. Our calculated modulation period for TW CrB is consistent with Applegate's findings but constraining the data to the more precise CCD and pe timings restricts this analysis to approximately 45% of one modulation period. Further timings during the predicted modulation period will be necessary to confirm this explanation. Equation (5) also makes the assumption that the modulation period of the binary is a constant. This is not necessarily the case when magnetic effects are present and variations in the binary modulation period may be observed.

The (O – C) residuals for the three approaches are drawn in Figure 10 for E≥0. The upper plot for the quadratic residuals is derived from Equation (2) which clearly shows the systematic error for epochs greater than zero. The middle plot is for episodic changes derived from Equation (4) and shows that some systematic errors remain. The bottom plot is for the sinusoidal residuals derived from Equation (5) and indicates that this is the best fit of the three m*odels* i.e. it's the closest to the zero residual horizontal line in the plot.

Another possible explanation for the O – C systematic errors of Figure 2 would be light-travel-time effects driven by the presence of a circumbinary sub-stellar companion. A similar analysis has been conducted by Kim, Jeong et al on YY Eridani, which is a W UMa binary.[18] Equation (5) would need to be modified to include the orbital parameters of the third body. Also more timings spanning the modulation period would be needed for a meaningful analysis to be undertaken on TW CrB.

A previous analysis of TW CrB has suggested that the period growth could be attributed to mass transfer between the components of the binary system at a rate of $2.74 \times 10^{-7}$ $M_\odot yr^{-1}$.[1] In this analysis Zhang & Zhang assumed that the primary component was a main sequence star of mass $1.19\ M_\odot$. Using the same assumption for primary star mass together the conservative mass transfer equation derived by Kwee ($\Delta P/P = 3(M_1/M_2 - 1)\Delta M_1/M_1$) and our calculated underlying change of orbital period of $6.66(14) \times 10^{-8}$ days $yr^{-1}$ and binary mass ratio 0.73,[19] we find an average mass transfer rate of $1.21(3) \times 10^{-7}$ $M_\odot yr^{-1}$.
This value is approximately half that estimated by Zhang & Zhang.

### *Light Curve Simulation*
The lightcurve simulation led to two interesting features being identified; a hotspot and two starspots.

The modelling process suggested a system that has two components each of which is nearly filling their Roche Lobe; that in turn supports the possibility of mass transfer with the generation of a hotspot. This is consistent with the work of other researchers: Zhang & Zhang, and Caballero-Nieves et al. It is also consistent with our findings of variation in period described elsewhere in this paper.

In addition we modelled two starspots to enhance the curve fit. These too are consistent with cyclic variations found in the period of the system and described elsewhere in this paper. A possible explanation for this is an Applegate type electromagnetic mechanism. The chromospheric activity implied by the starspots make this a very likely X-ray source. The ROSAT Bright Star Catalogue confirms that TW CrB is an X-ray source.[20] Also TW CrB is identified in SIMBAD with X-ray source 2XMM J160650.6 +271634[21]. However, our data does not allow us to investigate this further.

### Conclusions
We have calculated a new ephemeris for the near contact binary TW CrB based on all the available timings going back to 1946 and we have revised the average rate of change of the period of this system. During our investigation we found evidence that the period change is slowing and that the change may be cyclical, but there is insufficient ephemeris data to make a judgement on the mechanism causing this possible variation.
It is clear from the light curves and from the photometric solution that there is an increase in amplitude in the blue band with respect to the other two bands. This is likely to be caused by a hotspot on the primary companion which considering that the secondary has nearly filled its Roche lobe, implies evidence of mass transfer.

Clearly future research is needed to confirm the potential long-term cyclical behaviour of the period and thus be able to indicate the mechanism underpinning this behaviour. This would be a very long term project and it is hoped that investigators would take up this task in the future. Also radial velocity measurements would be needed to confirm the mass ratio and other parameters.

## Acknowledgements

We would like to thank Dr U. Kolb, Dr L. McComb and Dr F. Vincent of The Open University for their assistance in this work and the University for making the PIRATE observatory available as part of The Open University module S382, "Astrophysics". We extend our thanks to N. Cornwall, A. Grant, M. Hajducki, N. Smith and M. Treasure who assisted with some of the PIRATE observations. We would also like to thank Dr. D. Boyd, past President of the BAA, for his valuable suggestions and the referees for their constructive comments to enhance this paper.